\documentclass[final]{svjour3}
\usepackage[dvipdfmx]{graphicx}
\usepackage{rotating}
\usepackage{amssymb}
\usepackage{mathptmx}
\usepackage[numbers]{natbib}
\usepackage{xcolor}
\usepackage[caption=false]{subfig}
\makeatletter
\journalname{Journal of Low Temperature Physics}

\bibpunct{}{}{,}{s}{}{,}

\begin{document}

\newcommand{\hdblarrow}{H\makebox[0.9ex][l]{$\downdownarrows$}-}
\title{Noise temperature measurements for axion haloscope experiments at IBS/CAPP}

\author{S.W. Youn, E. Sala, J. Jeong, J. Kim and Y. K. Semertzidis}

\institute{Center for Axion and Precision Physics Research, Institute for Basic Science,\\ Daejeon, 34051, South Korea\\
\email{elena.sala@ibs.re.kr}}

\maketitle

\begin{abstract}

The axion was first introduced as a consequence of the Peccei-Quinn mechanism to solve the CP problem in strong interactions of particle physics and is a well motivated cold dark matter candidate.
This particle is expected to interact extremely weakly with matter and its mass is expected to lie in $\mu$eV range with the corresponding frequency roughly in GHz range. 
In 1983 P. Sikivie proposed a detection scheme, so called axion haloscope, where axions resonantly convert to photons in a tunable microwave cavity permeated by a strong magnetic field. 
A major source of the experimental noise is attributed to added noise by RF amplifiers, and thus precise understandings of amplifiers' noise is of importance. 
We present the measurements of noise temperatures of various low noise amplifiers broadly used for axion dark matter searches.


\end{abstract}

\section{Motivation}

The strong CP problem refers to the discrepancy between the experimental results suggest a conservation of the Charge-Parity symmetry in strong interactions and the predictions requiring a violating term in the Lagrangian \cite{1, 2, 3}. In 1977, R. Peccei and H. Quinn proposed a mechanism to solve the problem with the introduction of a global symmetry whose spontaneous breaking is associated with a Goldstone boson, named axion. The value of the axion mass is however arbitrary, but several ranges have been ruled out by cosmological constraints and astrophysical observations indicating its mass would fall in the $\mu$eV$-$meV range which corresponds to about $1-1000$\, GHz in frequency range. These properties make the axion an excellent cold dark matter candidate \cite{4}.

\subsection{Searching for axion at CAPP}
A fastly growing interest has brought several collaborations to search for axions. Among these, the Center for Axion and Precision Physics Research (CAPP) of the Institute for Basic Science (IBS) in South Korea was founded in 2013 aiming to contribute to axion physics with several experiments, most of them exploiting the haloscope technique. By now the center has built a facility equipped with dilution refrigerators and superconducting magnets which are used in parallel to search for axions in different mass ranges. At the same time R\&D projects are ongoing to enhance the sensitivity of the experiments by improving the cavity quality factor ($Q$) and cavity design in addition to developing low-noise high-gain amplifiers \cite{5}. 

\subsection{Axion haloscope and system noise}
In 1983, P. Sikivie proposed a method to detect axions based on the inverse Primakoff effect: using a resonant high $Q$ cavity permeated by a strong magnetic field, the axions are converted to microwave photons corresponding to the excitation quanta of the cavity mode \cite{6}. This is the working principle of haloscopes in which the conversion power on resonance is expected to be extremely weak: $\sim$ $10^{-21}$ W.
Since the axion mass is unknown, experiments need to scan a broad mass/frequency range as fast as possible; the scanning rate is thus a crucial quantity to represent the experimental sensitivity:%
\begin{equation}
	\frac{df}{dt} \propto \frac{B^{4}\,V^{2}C^{2}Q_{L}}{T_{\rm syst}^{2}},
	\label{eqn:sensitivity}
\end{equation}
where $B$ is the applied magnetic field, $V$ is the cavity volume, $C$ is the mode-dependent form factor, $Q_{L}$ is the loaded cavity quality factor and $T_{\rm syst}$ is the noise temperature of the system.\\
The sensitivity can be enhanced by incorporating high-field large-volume superconducting (SC) magnets and high $Q$ cavities with large volume and high form factors \cite{7, 8}. 

Another critical parameter which contributes to the sensitivity is the equivalent noise temperature of the system $T_{\rm syst}$. One of the main noise sources is due to the black body radiation emitted from the cavity, which can be mitigated by cooling the apparatus at ultra low temperatures ($\sim$mK) using for example dilution refrigerators. 
The other significant noise source is attributed to the electrical noise of the RF receiver chain, with the dominant contribution coming from the amplifier noise at the first stage, where a quantum-limited noise amplifier is usually placed to achieve the lowest possible noise level. A high electron mobility transistor (HEMT), widely used for axion experiments, is also one of the amplifiers that meet such requirements.
Even if its noise level is not quantum limited, due to its well-proven performance and stability, HEMTs are adopted for signal amplification for many axion experiments, such as those at CAPP. 
In addition, a detector chain, regardless of the type of amplifier at the first stage, i.e., quantum noise limited or not, consists of a series of amplifiers with a HEMT typically placed in the second stage and beyond.
In this regards, it is crucial to characterize their performance and precisely evaluate their noise figure (equivalently noise temperature) in the measurement setup.\\

\section{Method}
The thermal noise power generated by a perfectly matched resistor is given by:%
\begin{equation}
	P_{N} = kTB
	\label{eqn:power}
\end{equation}
where $k$ is the Boltzmann constant and $B$ is the bandwidth considered for the readout.
When the signal is amplified, the total readout noise has two contributions: the amplified thermal noise from the source and the noise generated by the amplifier itself. The most common method used to evaluate the noise temperature of the amplifiers relies on the Y-factor technique which consists in measuring two noise power levels from the device under test when connected to a noise source. The ratio between the device output noise power, $N$, with the source on and off is called the Y-factor \cite{9}.

The equivalent noise temperature $T_{e}$ of the device can be defined as %
\begin{equation}
	T_{e} = \frac{T_{ON} - YT_{OFF}}{Y - 1} \quad {\rm with} \quad Y = \frac{N_{ON}}{N_{OFF}} = \frac{T_{ON}}{T_{OFF}}
	\label{eqn:noiseT}
\end{equation}
where $T_{ON}$ and $T_{OFF}$ are the temperatures corresponding to the noise levels $N_{ON}$ and $N_{OFF}$.
Effectively, $T_{OFF}$ corresponds to the physical temperature of the noise source whereas $T_{ON}$ can be calculated from the Excess Noise Ratio (ENR) of the source:%
\begin{equation}
	{\rm ENR} = \frac{T_{ON} - T_{OFF}}{T_{0}}
	\label{eqn:ENR}
\end{equation}
where $T_{0}=290$ K is the reference temperature.\\
The noise measurement using the Y-factor technique consists of two steps: a calibration with just the noise source connected to a spectrum analyzer to measure the intrinsic noise of the receiver and a power measurement with the noise source connected to the input of the device.

\subsection{Measurements}
The goal of this work is to measure the noise temperature of the amplifiers used in the experimental set up in order to verify the expected noise level. Two different techniques were compared to evaluate the noise temperature of three HEMT amplifiers from LOW NOISE FACTORY, working in different frequency ranges. Both techniques were validated and the measured quantities were compared with the values provided by the supplier. The devices under test have working frequency ranges of 2-6 GHz, 4-8 GHz and 0.3-14 GHz. Since the actual experiment is performed at cryogenic temperature, the noise temperature of the devices was evaluated at 4 K assembling the amplifiers in a cryogenic unit already equipped with RF lines.\\
The measurements involve injection of noise signals with different power levels at the input of the amplifier. The output power spectra is evaluated using a spectrum analyser.
The first approach consisted in using a calibrated noise source with ENR of about 15 dB, whose dependency on frequency, in the range of interest, was extrapolated. The noise source was placed outside the cryostat and the measurements were performed both at room temperature and at cryogenic temperature allowing the validation of this method at 300 K and at 4 K. \\
In the second approach a heater was used as a noise source and connected to the amplifier placing a 40 dB attenuator between them. The attenuation is intended to ensure that the total integrated power at the input of the amplifier does not saturate the device. The voltage applied to the heater can be adjusted in order to set several temperatures, measured through diode sensors, and the spectrum from the amplifier was acquired for all the set points. This method was performed at 4 K and compared both with the measurements with the noise source and the specification of the amplifiers.\\

\subsection{Analysis}
To evaluate the noise temperature of the amplifiers using Eq. \ref{eqn:noiseT}, the noise power levels from the source have to be precisely evaluated. The attenuation of RF lines, attenuators and other passive components reduces the power level and thus the effect should also be integrated in the analysis. The attenuation due to the input and output lines was measured terminating the end of the cables with 50 Ohm and recording the cable reflection with a network analyzer. The attenuation has a dependence upon the frequency: the maximum measured value is about 4 dB and 7 dB at room and low temperature respectively.\\
Considering the room temperature measurement with the noise source, the actual temperature $T$ at the amplifier input is evaluated using the following formula:%
\begin{equation}
T = T_{ON} \times A + (1 - A)T_{0},
\label{eqn:attenuation}
\end{equation}
with the noise source nominal value, $T_{ON}$ in Eq. \ref{eqn:ENR}, and the input line attenuation $A$.
In this calculation the frequency dependence of the noise source ENR and the line attenuation were taken into account.\\
For the low temperature measurements using the noise source, the line connecting the source to the amplifier should be characterized by considering not only the attenuation but also the temperature gradient, which affects the noise power. To evaluate the temperature corresponding to the noise power level at the amplifier input, the physical temperature of the cryostat stages and of the passive components inserted on the line was measured by diode sensors. Eq. \ref{eqn:attenuation} is thus modified to evaluate the attenuation of the noise power from the source at each stage of the cryostat until the 30 dB attenuator placed before the amplifier. This component is needed to avoid the saturation of the device; the noise signal arriving to the amplifier is then evaluated considering the 30 dB attenuation using Eq. \ref{eqn:attenuation}.\\
For both techniques the noise source power level at the input of the amplifier was thus corrected for the variation of its physical temperature due to the line attenuation, temperature gradient and passive components effects.\\
The measurements with a heat load used as a noise source do not involve any input line since the load is directly connected to the amplifier through a 40 dB attenuator placed at the input. This method allows to set several temperatures varying the voltage across the heater; the noise power corresponding to those temperatures is attenuated by the attenuator and the spectra from the amplifier acquired for various voltages. Considering Eq. \ref{eqn:power}, the noise temperature is evaluated through a linear fit of the recorded noise levels as a function of the input power.\\

\begin{figure}[h]
\begin{center}
       { \includegraphics[width=.32\linewidth]{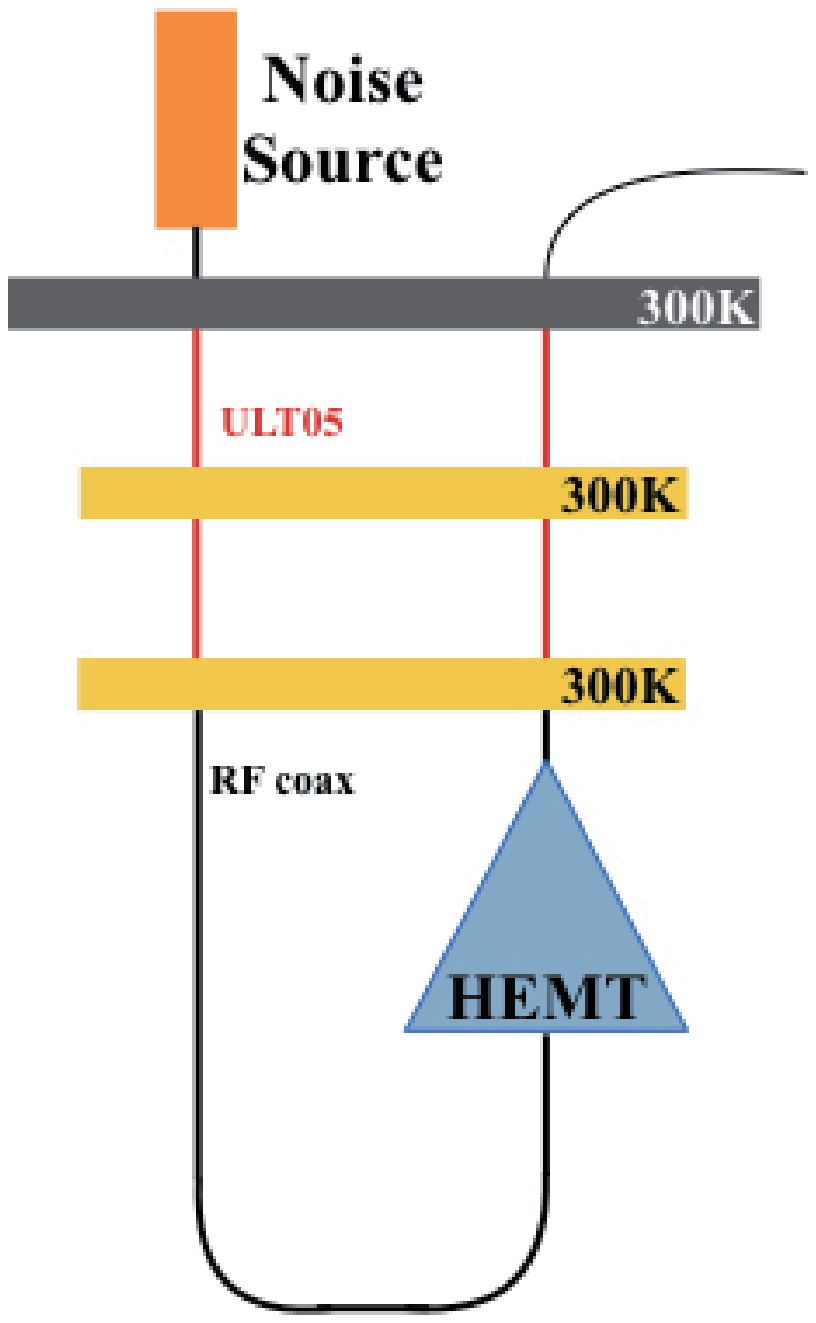}
    }\hfill
       { \includegraphics[width=.32\linewidth]{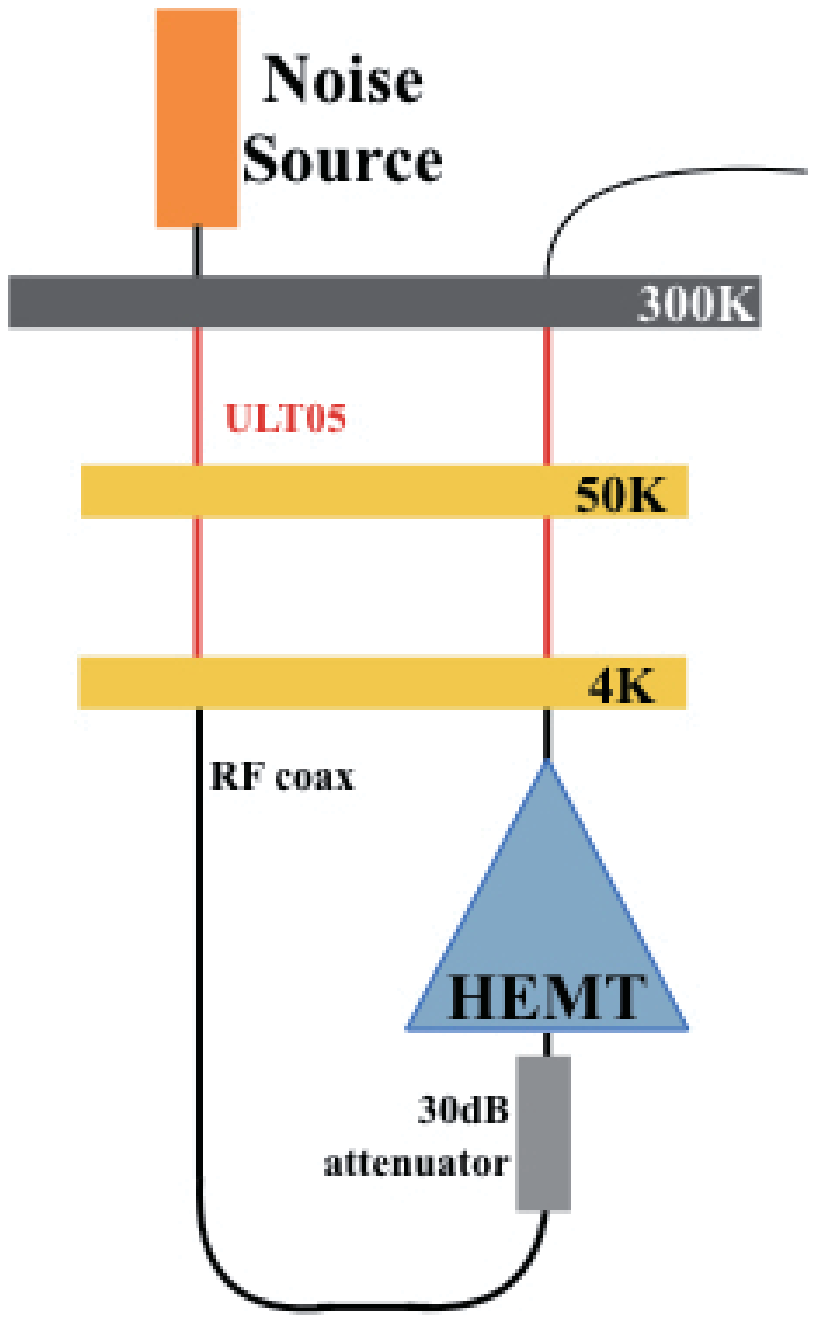}
    }\hfill
       { \includegraphics[width=.32\linewidth]{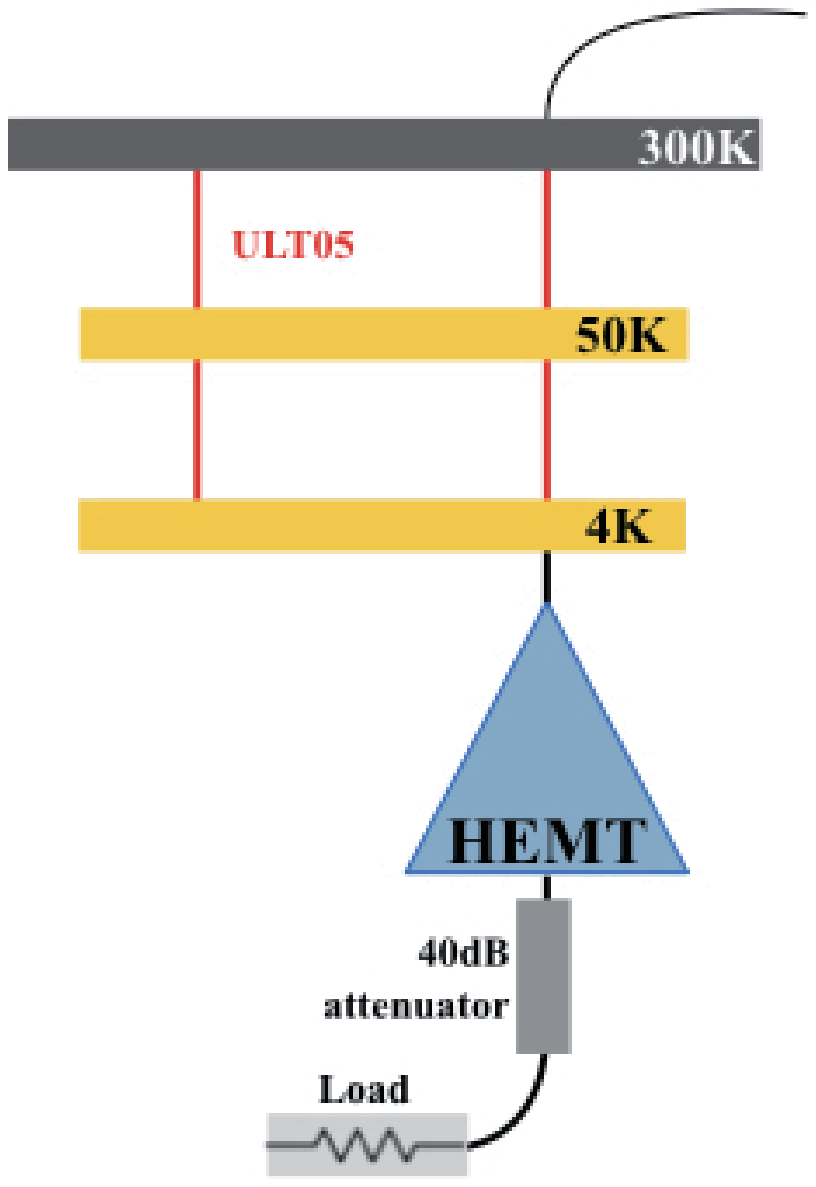}
    }\\
    \caption{Measurement configurations. {\it Left}: Room temperature measurement with a calibrated noise source. {\it Center}: Low temperature measurement with the calibrated noise source. {\it Right}: Low temperature measurement using a load. }
    \label{fig:conf}
    \end{center}
\end{figure}

\section{Results and conclusion}
\subsection{Room temperature measurements with a noise source}
\begin{figure}[h]
\begin{center}
       { \includegraphics[width=.32\linewidth]{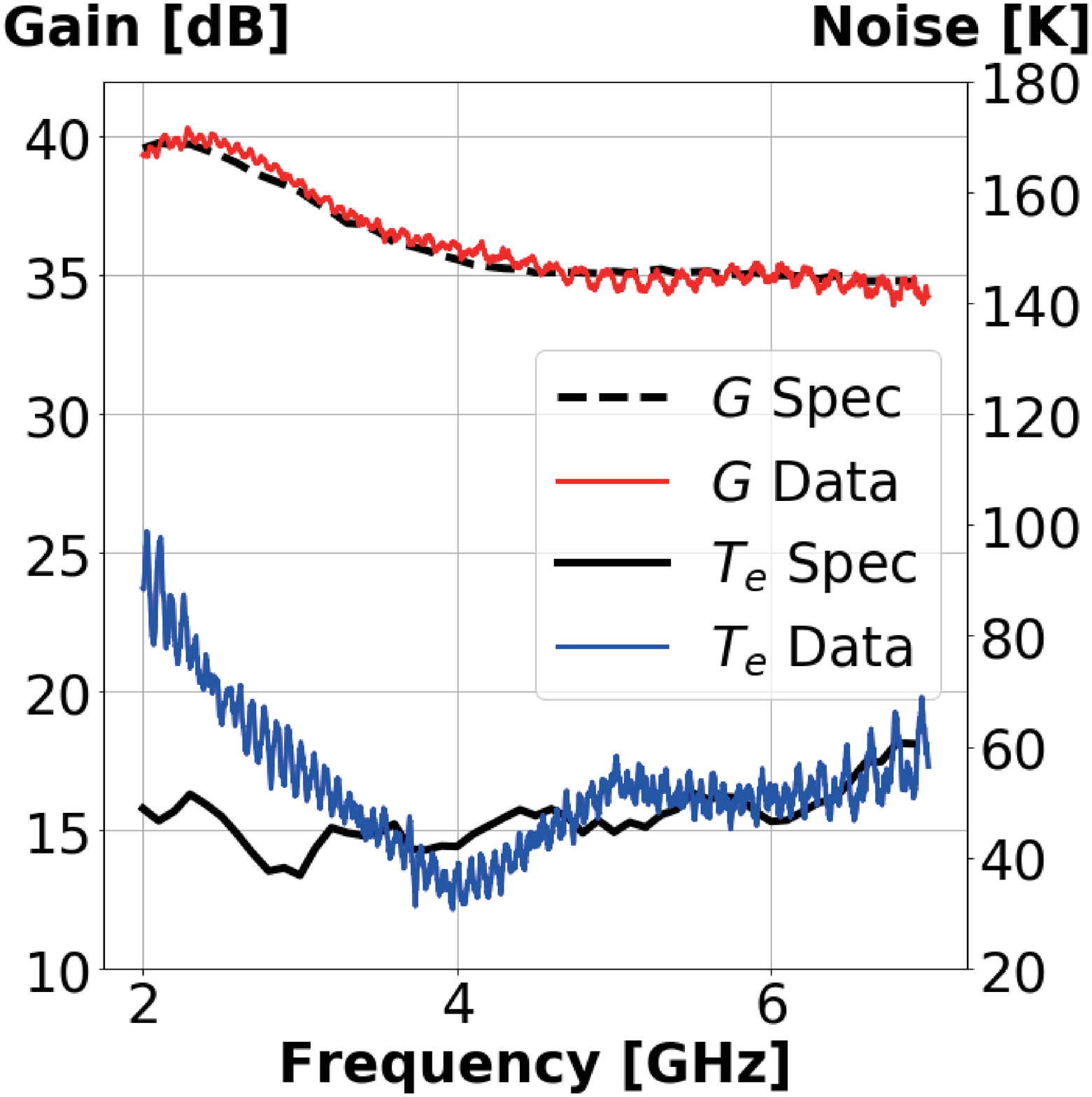}
    }
     {   \includegraphics[width=.32\linewidth]{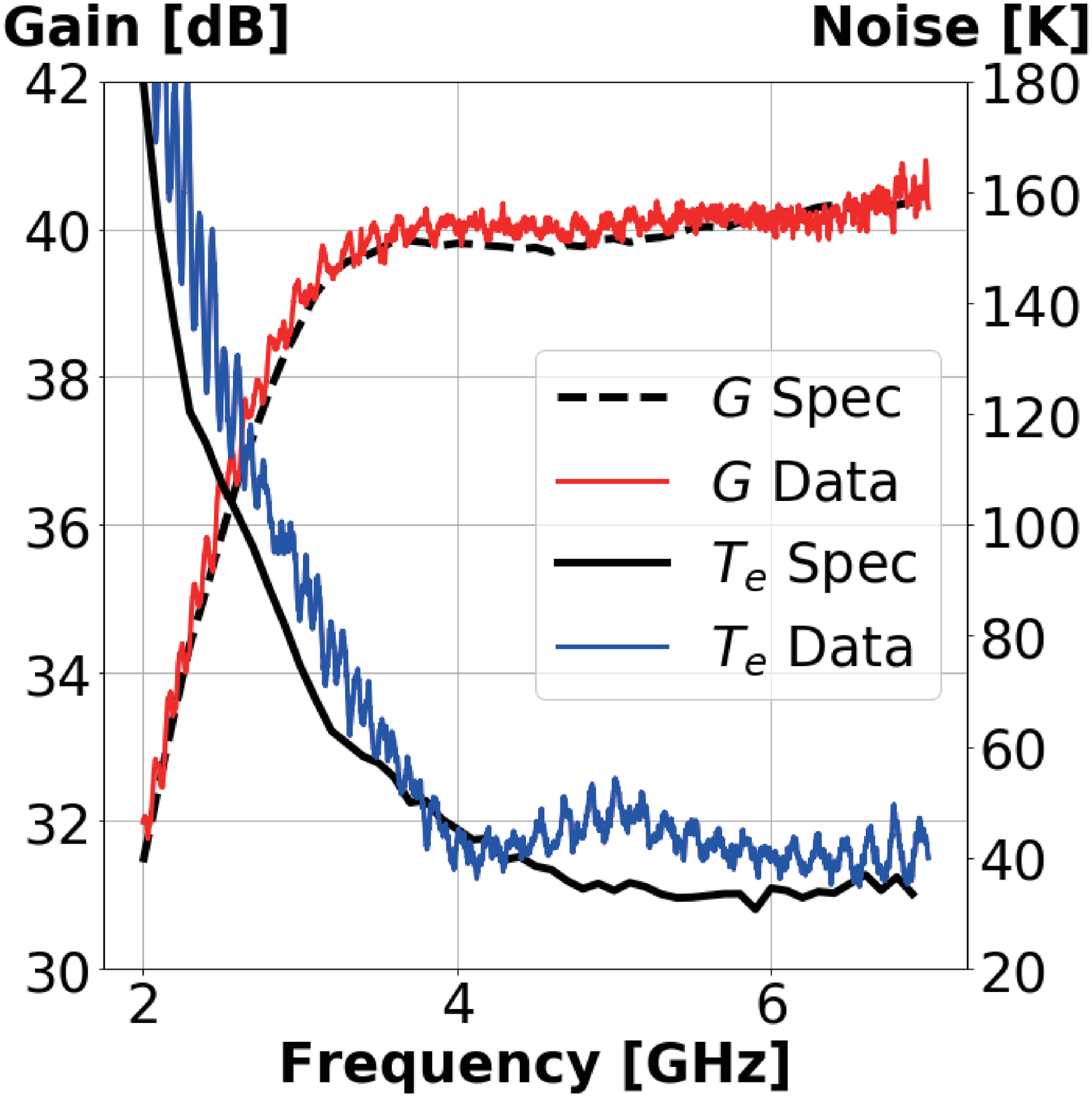}
    }
     {   \includegraphics[width=.32\linewidth]{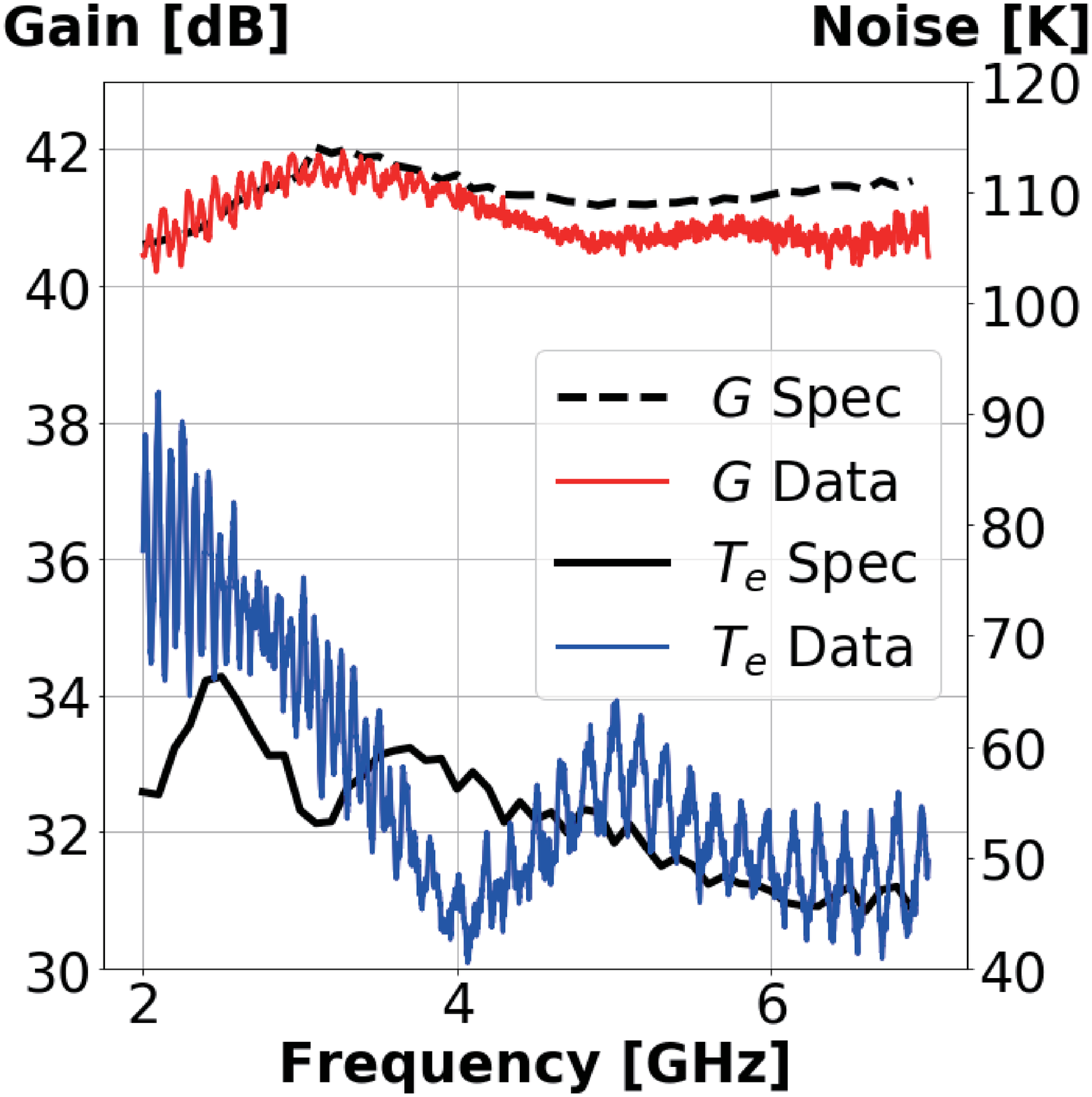}
    }
    \caption{Room temperature measurements, T = 300 K: comparison between the gain ($G$) and noise temperature ($T_{e}$) measurements data and the reference values form the supplier. {\it Left}: 2 to 6 GHz range HEMT amplifier. {\it Center}: 4 to 8 GHz range HEMT amplifier. {\it Right}: 0.3 to 14 GHz range HEMT amplifier. }
    \label{fig:room}
   \end{center}
\end{figure}

The first measurement consisted in the validation of the Y-factor method using the calibrated 15 dB ENR noise source at room temperature. To assure that this measurement could be compared with the low temperature one, the same measurement set up was used  (Fig.~\ref{fig:conf}): the noise source was placed outside the cryostat on the 300 K stage and connected to the amplifier placed on the 4 K plate of the cryostat through the available RF line. Two spectra, one for each amplifier, were acquired corresponding to the power levels with the noise source off and on. The gain and the noise temperature were then obtained using the Y-factor method and the results compared to the values provided by the supplier for a measurement at 300 K. The comparison shows a good agreement in this configuration for all the tested amplifiers in their working ranges, Fig.~\ref{fig:room}.

\subsection{Low temperature measurements with noise source}
\begin{figure}[h]
\begin{center}
     {   \includegraphics[width=.32\linewidth]{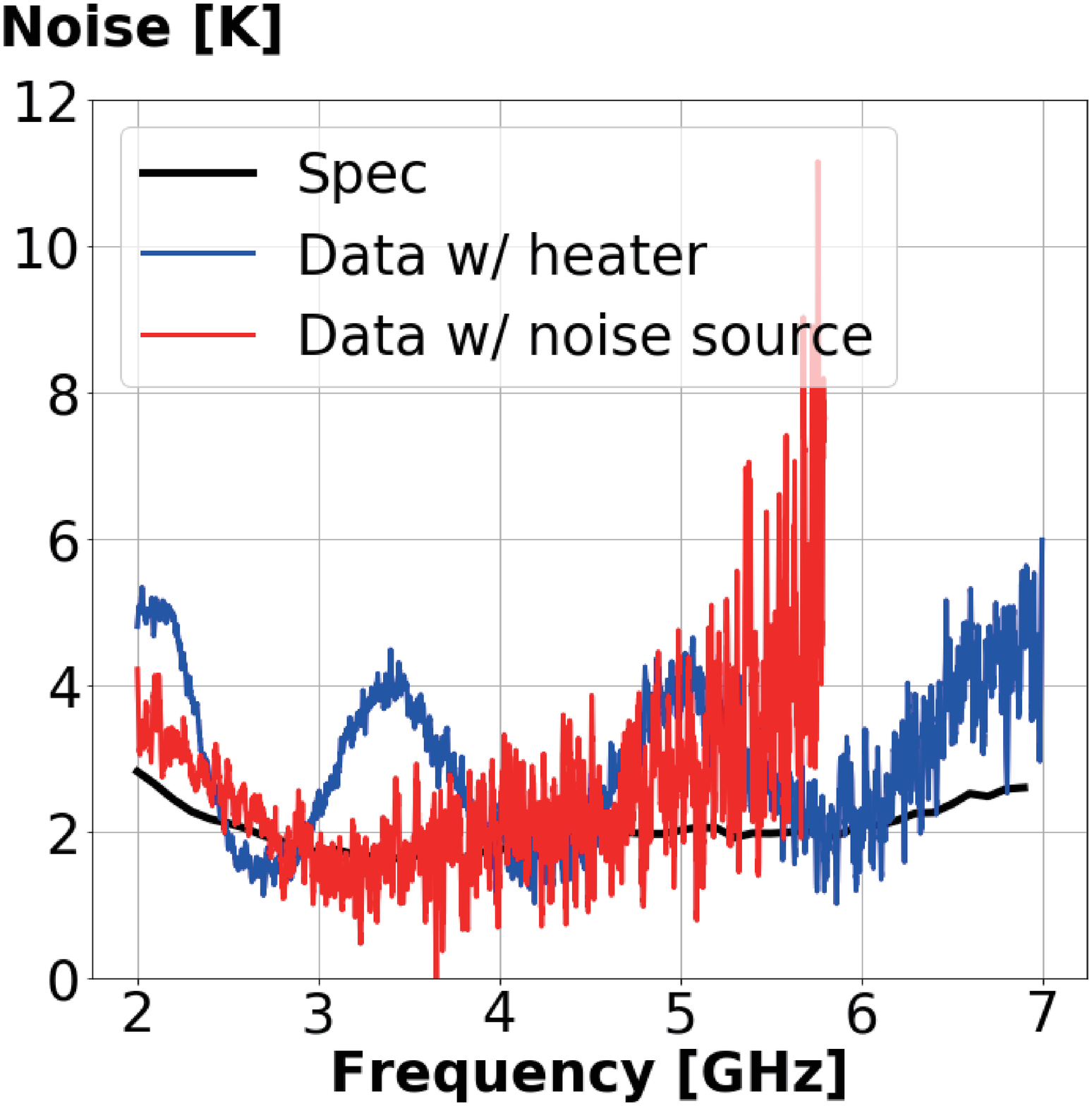}
    }\hfill
      {  \includegraphics[width=.32\linewidth]{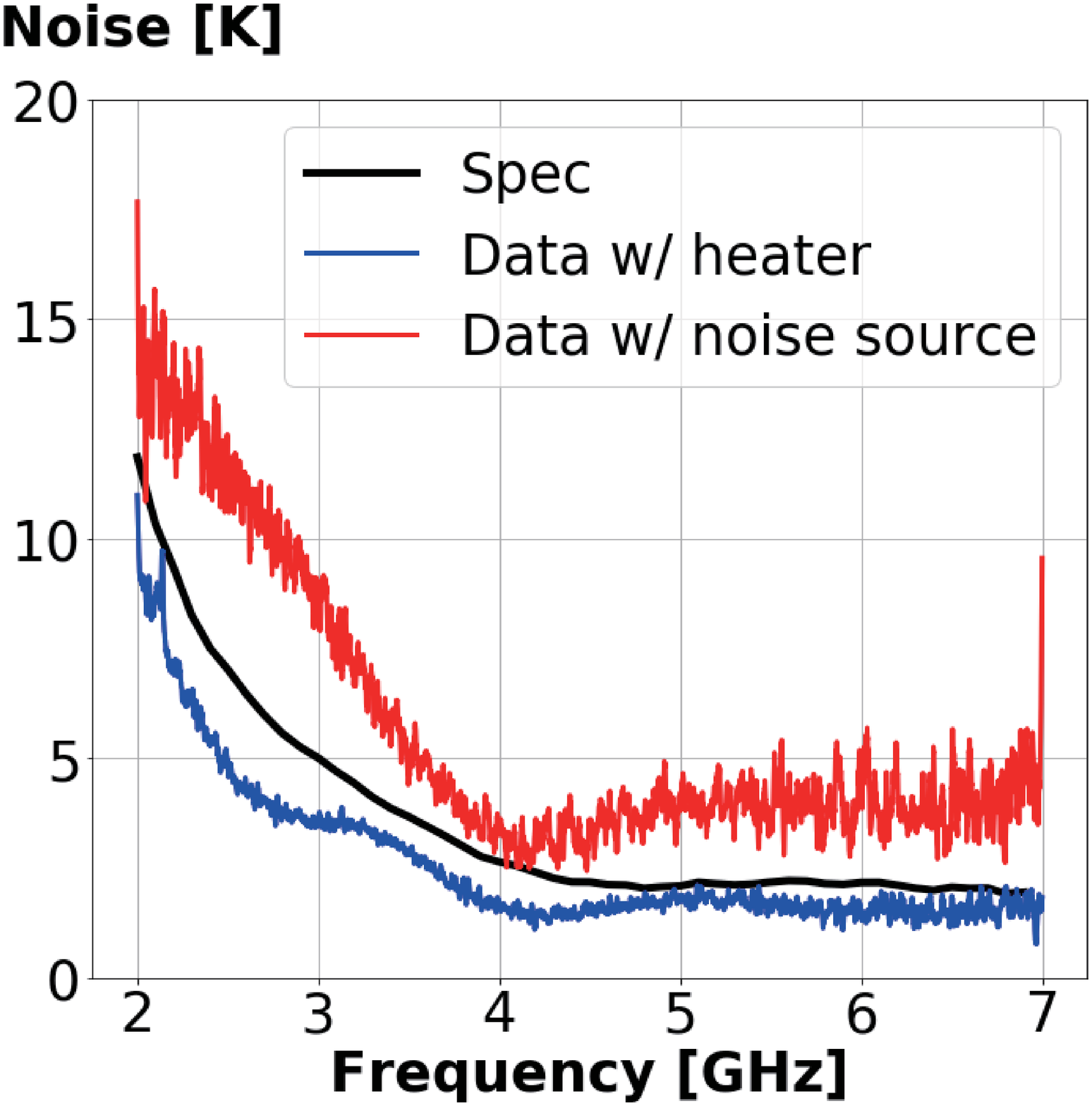}%
    }\hfill
     {   \includegraphics[width=.32\linewidth]{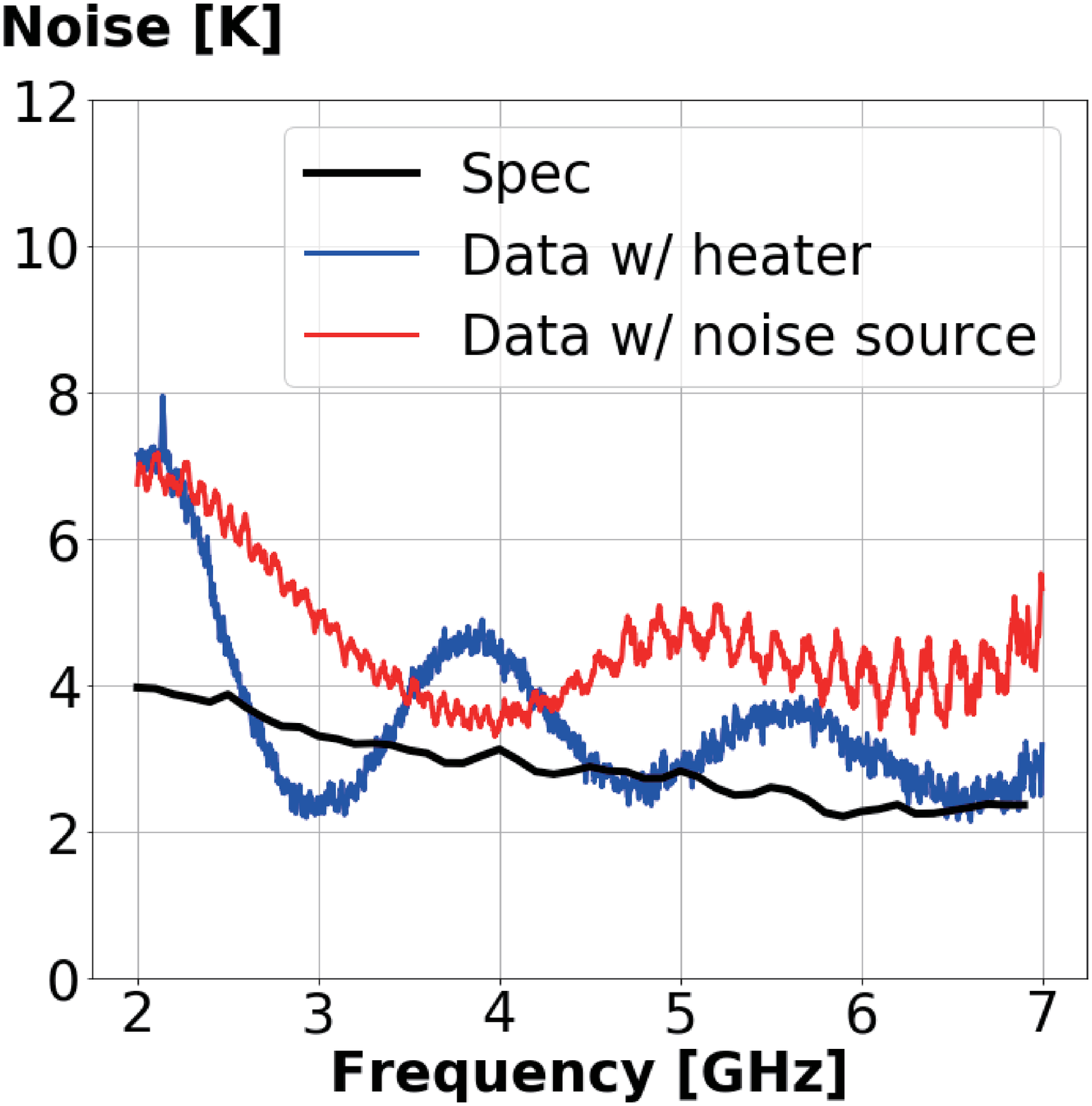}%
    }\\
    \caption{Low temperature measurements, T $\sim$ 5 K: comparison between the noise temperature ($T_{e}$) measurements data with the heater (blue line), with the noise source (red line) and the reference values form the supplier (black line). {\it Left}: 2 to 6 GHz range HEMT amplifier. {\it Center}: 4 to 8 GHz range HEMT amplifier. {\it Right}: 0.3 to 14 GHz range HEMT amplifier.}
    \label{fig:LT}
    \end{center}
\end{figure}

The second measurement, with the noise source at 300 K, was performed by cooling the apparatus at about 4 K; a 40 dB attenuator was placed at the input of the amplifier to guarantee that the noise power would not excessively exceed the noise temperature level of the amplifier and to make the temperature gradient negligible. Both gain and noise temperature were evaluated and found consistent with the manufacturer values for 5 K temperature measurement: the comparison is shown in Fig.~\ref{fig:LT}.

\subsection{Low temperature measurements using a heat load}
The last set of measurements was performed using a heater connected to a 40 dB attenuator at the input of the amplifier to reach the appropriate noise power level; the configuration is shown in Fig.~\ref{fig:conf}.
The output spectrum of the amplifier was recorded for several points, corresponding to the voltage$/$temperature of the heater, ensuring that the set value was stable. The results although the presence of a mismatch that produce the wiggling in the plots, are in good agreement with the measurements with the noise source and the specification of the amplifiers, Fig.~\ref{fig:LT}.

\subsection{Conclusions}
The noise temperature and the gain of some HEMT amplifiers typically used for axion search experiments were evaluated using the Y-factor method involving a noise source and a load. The method was first verified at room temperature reaching a good agreement between the evaluated gain and noise temperature parameters and those specified by the amplifier producer. The measurement was then performed at low temperature with both noise sources; comparing the results it was possible to conclude that both techniques can provide the expected results given by the supplier. Being the noise temperature an important parameter that can affect the sensitivity of the experiment, it is mandatory to understand it with the best possible precision. These techniques can thus be implemented in the experimental set up to ensure the evaluation of the actual noise level. The next step will thus be the evaluation of the noise level due to the entire electronic chain.

\begin{acknowledgements}
This work was supported by the Institute for Basic Science (IBS-R017-D1-2019-a00/IBS-R017-Y1-2019-a00)
\end{acknowledgements}


\end{document}